\documentclass[twocolumn,showpacs]{revtex4}
\usepackage{epsf,graphicx}
\usepackage{subfigure}
\usepackage{amsmath,amssymb,amsfonts}
\begin{document}
\newcommand{\avg}[1]{\left\langle{#1}\right\rangle}
\newcommand{\davg}[1]{\left\langle\!\left\langle{#1}\right\rangle\!\right\rangle}
\newcommand{\ovl}[1]{\overline{#1}}
\renewcommand{\l}{\left}
\renewcommand{\r}{\right}

\author{E.V.~Votyakov{$^{1*}$}, H.I.~Hidmi{$^2$}, A.~De~Martino{$^{1\dag}$} and D.H.E.~Gross{$^{1\ddag}$}}

\affiliation{{$^1$}Hahn-Meitner-Institut, Bereich Theoretische
Physik, Glienickerstr. 100, 14109 Berlin (Germany)\\
{$^2$}Bethlehem University, PO Box 09, Bethlehem (Palestine)}

\title{\sffamily Microcanonical mean-field thermodynamics of
self-gravitating and rotating systems}


\begin{abstract}
We derive the global phase diagram of a self-gravitating $N$-body
system enclosed in a finite three-dimensional spherical volume $V$
as a function of total energy and angular momentum, employing a
microcanonical mean-field approach. At low angular momenta (i.e.
for slowly rotating systems) the known collapse from a gas cloud
to a single dense cluster is recovered. At high angular momenta,
instead, rotational symmetry can be spontaneously broken and
rotationally asymmetric structures (double clusters) appear.
\end{abstract}

\maketitle

The statistical equilibrium properties of systems of particles
interacting via long-range forces (the so-called non-extensive
systems) are currently the subject of intense research, both for
their highly non-trivial thermodynamics (displaying such features
as negative heat capacities \cite{thirring70,gross174}) and for
the considerable conceptual and technical difficulties they
present. It is known that the long-range nature of the potential
makes the canonical ensemble inadequate for describing their
statics \cite{landau96,gallavotti99,padmanabhan90}, because the
usual thermodynamic limit, where (number of particles)
$N\to\infty$ and (volume) $V\to\infty$ while intensive variables
are kept fixed, does not exist. A central issue is hence whether
phase transitions and other conventional statistical phenomena are
possible in non-extensive systems \cite{gross186}.

Among non-extensive systems, self-gravitating gases, i.e. systems
of classical particles subject to mutual gravitation, have
deserved the most attention. Their usual static description is
based on the microcanonical ensemble \cite{padmanabhan90}. In this
framework, the key problem is finding the most probable
equilibrium configuration of a self-gravitating gas enclosed in a
finite $3$-dimensional box of volume $V$ as a function of the
conserved quantities (integrals of motion), the simplest (but
possibly not the only relevant ones) being the total energy $E$
and the total angular momentum $\boldsymbol{L}$.

Dynamical methods \cite{chandrasekhar69} based on fluid-mechanics
techniques suggest (see e.g. \cite{lyndenbell66}) that upon
increasing the ratio between rotational and gravitational energy,
the stationary distribution can change from a single dense cluster
to a double cluster, and that other structures such as disks and
rings might appear.

On the other hand, so far static theories could not recover the
richness of the dynamical picture. Taking the total energy as the
only control parameter (see \cite{padmanabhan90} for a review and
\cite{chavanis01,pettini01,chavanis02} for more recent work and
references) after removing the rotational symmetry artificially
e.g. by constraining the system into a non-spherical box, a
``collapse'' transition has been found \cite{thirring70}, where,
as the energy (temperature) is lowered, the equilibrium
configuration changes from a homogeneous cloud to a dense cluster
lying in an almost void background, with an intermediate
``transition'' regime characterized by negative specific heat.
Despite some attempts \cite{gross181}, a detailed static theory
embodying angular momentum is lacking.

In this work, building essentially on
\cite{lyndenbell66,laliena98}, we aim at bridging the gap between
the static and the dynamical approaches by analyzing the statics
of a self-gravitating and rotating gas in a microcanonical setting
including angular momentum. Within a mean-field approximation, we
derive an algebraic integral equation for the density profiles
maximizing the microcanonical entropy and solve it numerically as
a function of $E$ and $L$. Along with the usual collapse,
occurring at low $L$, we find that for sufficiently high $L$ the
rotational symmetry of the Hamiltonian can be spontaneously
broken, giving rise to more complex equilibrium distributions,
including double clusters, rings and disks. The global phase
diagram of the system is presented. We shall concentrate here on
the equilibrium density profiles, deferring a detailed discussion
of the related (highly non-trivial) thermodynamic picture to a
more extensive report.

We consider the $N$-body system with Hamiltonian
\begin{equation} \label{ham} H_N\equiv
H_N(\{\boldsymbol{r}_i\},\{\boldsymbol{p}_i\})=\frac{1}{2m}
\sum_{i=1}^N p_i^2+\Phi(\{\boldsymbol{r}_i\})
\end{equation}
with $\Phi(\{\boldsymbol{r}_i\})=-Gm^2\sum_{i<j}
|\boldsymbol{r}_i-\boldsymbol{r}_j|^{-1}$. $\boldsymbol{r}_i$,
$\boldsymbol{p}_i$ and $m$ denote, respectively, the position,
momentum and mass of the $i$-th particle. The system is assumed to
be enclosed in a spherical volume $V$ (to preserve rotational
symmetry and ensure angular momentum conservation). The crucial
quantity to be evaluated is the microcanonical ``partition sum''
\begin{equation}\label{wu}
W_N(E,\boldsymbol{L})\!=\!\frac{\epsilon}{N!}\!\int\!\!\delta(H_N-E)
~\delta(\boldsymbol{L}-\sum_{i=1}^N\boldsymbol{r}_i
\times\boldsymbol{p}_i)D\boldsymbol{r}D\boldsymbol{p}
\end{equation}
where $D\boldsymbol{r}=\prod_{i=1}^Nd\boldsymbol{r}_i$,
$D\boldsymbol{p}=\prod_{i=1}^N(d\boldsymbol{p}_i/h^3)$, and
$\epsilon$ is a constant that makes $W_N$ dimensionless. According
to Boltzmann, the entropy is given by
\begin{equation}\label{duino}
S_N(E,\boldsymbol{L})=\ln W_N(E,\boldsymbol{L})
\end{equation}
Our aim is to find the density profiles that maximize $S_N$.

Following Laliena \cite{laliena98}, the integration over momenta
in (\ref{wu}) can be carried out using a Laplace transform. This
yields
\begin{equation}\label{wu2}
W_N(E,\boldsymbol{L})\!=\!\frac{A}{N!}\!\int\![E-\frac{1}{2}\boldsymbol{L}^T
\mathbb{I}^{-1}\boldsymbol{L}-\Phi(\{\boldsymbol{r}_i\})]^{\frac{3N-5}{2}}
D\boldsymbol{r}
\end{equation}
where $A$ is a constant and
$\mathbb{I}\equiv\mathbb{I}(\{\boldsymbol{r}_i\})$ is the inertia
tensor, with elements $I_{ab}(\{\boldsymbol{r}_i\})=m\sum_{i=1}^N
(r_i^2\delta_{ab}-r_{i,a} r_{i,b})$ ($a,b=1,2,3$).

In order to evaluate the integral over $V^N$, we use the following
mean field approximation. Letting $\rho(\boldsymbol{r})$ denote
the particles' density inside $V$
($\int\rho(\boldsymbol{r})d\boldsymbol{r}=N$), we set
\begin{gather}
\Phi(\{\boldsymbol{r}_i\})\to
\Phi[\rho]=-\frac{Gm^2}{2}\int\frac{\rho(\boldsymbol{r})\rho(\boldsymbol{r}')}{
|\boldsymbol{r}-\boldsymbol{r}'|}d\boldsymbol{r}d\boldsymbol{r'}\\
I_{ab}(\{\boldsymbol{r}_i\})\to I_{ab}[\rho]=m\int
\rho(\boldsymbol{r})\l(r^2\delta_{ab}-r_a r_b\r)d\boldsymbol{r}
\end{gather}
so that (\ref{wu2}) can be re-cast in the form of the
functional-integral
\begin{equation}\label{wu3}
W_N(E,\boldsymbol{L})\!=\!\frac{A}{N!}\!\int\![E-\frac{1}{2}\boldsymbol{L}^T
\mathbb{I}^{-1}\boldsymbol{L}-\Phi[\rho]]^{\frac{3N-5}{2}}\!P[\rho]d\rho(\boldsymbol{r})
\end{equation}
where $P[\rho]$ is the probability to observe a density profile
$\rho\equiv \rho(\boldsymbol{r})$. To estimate $P[\rho]$, we adopt
the logic of Lynden-Bell \cite{lyndenbell66}. We subdivide the
spherical volume $V$ into $K$ identical cells labeled by the
position of their centers. The idea is to replace the integral
over $V$ with a sum over the cells. In order to avoid overlapping
we assume that each cell may host up to $n_0$ particles ($1\ll
n_0\ll N$). This condition is essentially equivalent to
introducing a hard-core for each particle, and projects out all
the physics that is expected to play a role at short distances.
$P[\rho]$ is now proportional to the number of ways in which our
$N$ particles can be distributed inside the $K$ cells with maximal
capacity $n_0$. Denoting by $n(\boldsymbol{r}_k)$ the number of
particles located inside the $k$-th cell, a simple combinatorial
reasoning \cite{lyndenbell66} leads to
\begin{equation}
%P[\rho]\propto\frac{N!}{n(\boldsymbol{r}_1)!\cdots
%n(\boldsymbol{r}_K)!}\prod_{\text{cells }
%}\frac{n_0!}{(n_0-n(\boldsymbol{r}_k))!}
P[\rho]\propto N!\prod_{\text{cells }
k}\binom{n_0}{n(\boldsymbol{r}_k)}
\end{equation}
Defining the relative cell occupancy
$c(\boldsymbol{r})=n(\boldsymbol{r})/n_0$, so that
$\rho(\boldsymbol{r})=K n_0 c(\boldsymbol{r})/V$, and applying
Stirling's formula to approximate the factorials (assuming
$n(\boldsymbol{r}_k)\gg 1$ and $n_0-n(\boldsymbol{r}_k)\gg 1$),
one obtains
\begin{multline}
P[c]\propto  N!~e^{-\frac{n_0 K}{V}\int \l[c(\boldsymbol{r})\log
c(\boldsymbol{r})+
(1-c(\boldsymbol{r}))\log(1-c(\boldsymbol{r}))\r]
d\boldsymbol{r}}=\\
  = N!~ e^{-\frac{N}{\Theta}\int
 \l[c(\boldsymbol{x})\log c(\boldsymbol{x})+
(1-c(\boldsymbol{x}))\log(1-c(\boldsymbol{x}))\r] d\boldsymbol{x}}
\end{multline}
where we introduced the dimensionless variable $x=r/R$ and defined
$\Theta=\frac{NV}{n_0 K R^3}=\int
c(\boldsymbol{x})d\boldsymbol{x}$.

Plugging this expression into (\ref{wu3}) one easily arrives at
\begin{equation}\label{entra}
W_N(E,\boldsymbol{L})\equiv e^{S_N(E,\boldsymbol{L})}\propto\int
e^{N\l(s_1[c]+s_2[c]\r)}~dc(\boldsymbol{x})
\end{equation}
where $s_1$ and $s_2$ are given by
\begin{gather}
s_1[c]=\frac{3}{2}\log[E-\frac{1}{2}\boldsymbol{L}^T
\mathbb{I}^{-1}\boldsymbol{L}-\Phi[c]]\\s_2[c]\!=
\!-\frac{1}{\Theta}\!\int\![c(\boldsymbol{x})\log
c(\boldsymbol{x}) +(1-c(\boldsymbol{x}))\log(1-c(\boldsymbol{x}))]
d\boldsymbol{x}\nonumber
\end{gather}
Notice that $\mathbb{I}\equiv\mathbb{I}[c]$. We have neglected
terms appearing in the exponent which do not scale with $N$.

For large $N$ one can resort to the saddle-point method to
evaluate (\ref{entra}). Variation of the entropy $S_N$ with
respect to the relative cell occupancy $c$, with the constraint on
$\Theta$ enforced by a Lagrange multiplier $\mu$, leads to the
stationarity condition
\begin{equation}\label{key}
\log\frac{c(\boldsymbol{x})}{1-c(\boldsymbol{x})}=
-\frac{\beta}{\Theta}U(\boldsymbol{x})+\frac{1}{2}\beta(
\boldsymbol{\omega}\times\boldsymbol{x})^2-\mu
\end{equation}
or, equivalently,
\begin{equation}\label{key2}
c(\boldsymbol{x})=(1+e^{\frac{\beta}{\Theta}U(\boldsymbol{x})-\frac{1}{2}\beta(
\boldsymbol{\omega}\times\boldsymbol{x})^2+\mu})^{-1}
\end{equation}
where $\boldsymbol{\omega}$ is the angular velocity (related to
the total angular momentum by the relation
$\boldsymbol{L}=\mathbb{I}\boldsymbol{\omega}$), and the
shorthands $\beta$ and $U(\boldsymbol{x})$ are respectively
defined as
\begin{gather}
\beta=\frac{3/2}{[E-\frac{1}{2}\boldsymbol{L}^T
\mathbb{I}^{-1}\boldsymbol{L}-\Phi[c]]}\\
U(\boldsymbol{x})=-\int\frac{c(\boldsymbol{x'})}{
|\boldsymbol{x}-\boldsymbol{x'}|}~d\boldsymbol{x'}
\end{gather}
The essence of our mean-field approach becomes clear if we notice
that $\Phi[c]\propto\int
c(\boldsymbol{x})U(\boldsymbol{x})d\boldsymbol{x}$.

Eqs (\ref{key}) or (\ref{key2}) are our central result. They hold
for any long-range potential $\Phi$ for which our mean-field
approximation can be justified. Functions $c(\boldsymbol{x})$
solving them and corresponding to entropy maxima represent our
desired equilibrium distribution of particles. Of course, for
fixed energy and angular momentum there may exist different
solutions, each having its entropy. In such cases, the higher the
entropy the more probable the solution.

Technically, one can only hope to solve e.g. (\ref{key2}) by
numerical integration. However, the implicit dependence of
$U(\boldsymbol{x})$ on $c(\boldsymbol{x})$ via a 3-dim. integral
makes this a formidable task. To simplify things, we pass to
spherical coordinates, $\boldsymbol{x}=(x,\theta,\phi)$, and
expand our potential and relative occupancy in series of real
spherical harmonics:
\begin{gather}
\frac{1}{|\boldsymbol{x}-\boldsymbol{x'}|}\!=\!\sum_{l=0}^\infty
\!\sum_{m=-l}^l\!\frac{4\pi}{2l+1}\frac{(x\vee x')^l}{(x\wedge
x')^{l+1}}Y_{lm}(\theta,\phi)Y_{lm}(\theta',\phi')\nonumber\\
c(\boldsymbol{x})=\sum_{l=0}^\infty \sum_{m=-l}^l
b_{lm}(x)Y_{lm}(\theta,\phi)\label{ser}
\end{gather}
with $x\vee x'=\min\{x,x'\}$ and $x\wedge x'=\max\{x,x'\}$.
$b_{lm}(x)$ is a radial function whose precise form we shall soon
derive. Using the series (\ref{ser}), together with the
completeness relation for our basis set $\{Y_{lm}\}$, one can
easily show that
$U(\boldsymbol{x})=\sum_{l,m}u_{lm}(x)Y_{lm}(\theta,\phi)$, with
\begin{equation}\label{u}
u_{lm}(x)=-\frac{4\pi}{2l+1}\int\frac{(x\vee x')^l}{(x\wedge
x')^{l+1}} b_{lm}(x')(x')^2 dx'
\end{equation}
Multiplying both sides of (\ref{key2}) by $Y_{lm}$ and integrating
over angular variables one finds
\begin{widetext}\vspace*{-0.5cm}
\begin{equation}\label{due}
b_{lm}(x)=\int Y_{lm}(\theta,\phi)\l[1+e^{\frac{\beta}{\Theta}
\sum_{l=0}^\infty \sum_{m=-l}^l u_{lm}(x)Y_{lm}(\theta,\phi)
-\frac{1}{2}\beta\omega^2 x^2 \sin^2\theta+\mu}\r]^{-1}\sin\theta
d\theta d\phi
\end{equation}
\end{widetext}\vspace*{-0.7cm}
where $l=0,1,\ldots$ and $m=-l,-l+1,\ldots,l$.
\begin{figure}
\includegraphics[width=8.5cm,height=5.3cm,clip=true]{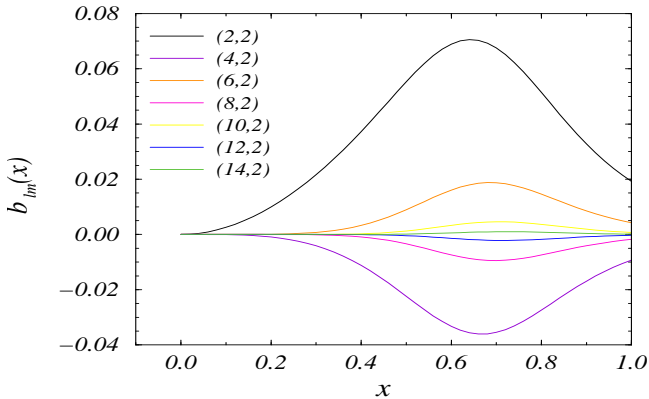}
\caption{\label{spag}Typical behaviour of the radial part $b_{lm}$
as a function of $l$ for $m=2$. This plot was obtained for
$E=-0.18$ and $L=0.44$.}
\end{figure}
\begin{figure}
\includegraphics*[width=8.5cm,height=5.3cm,clip=true]{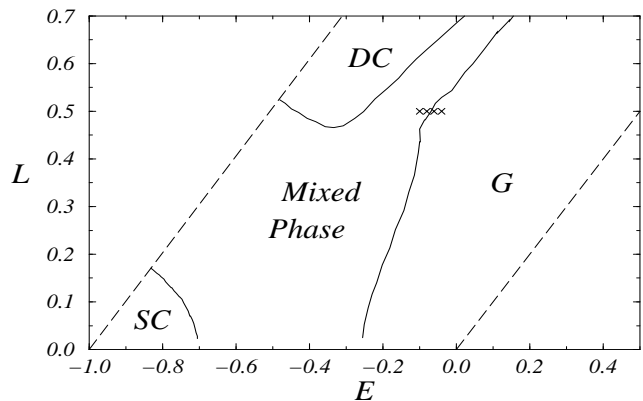}
\caption{\label{phased}Global phase diagram of the
self-gravitating $N$-body system with angular momentum. The dashed
lines delimit the region where $\mathcal{H}[S_N]$ was computed.}
\end{figure}
\begin{figure}[!]
\subfigure[$E=-0.72$,
$L=0.4$]{\scalebox{.49}{\includegraphics{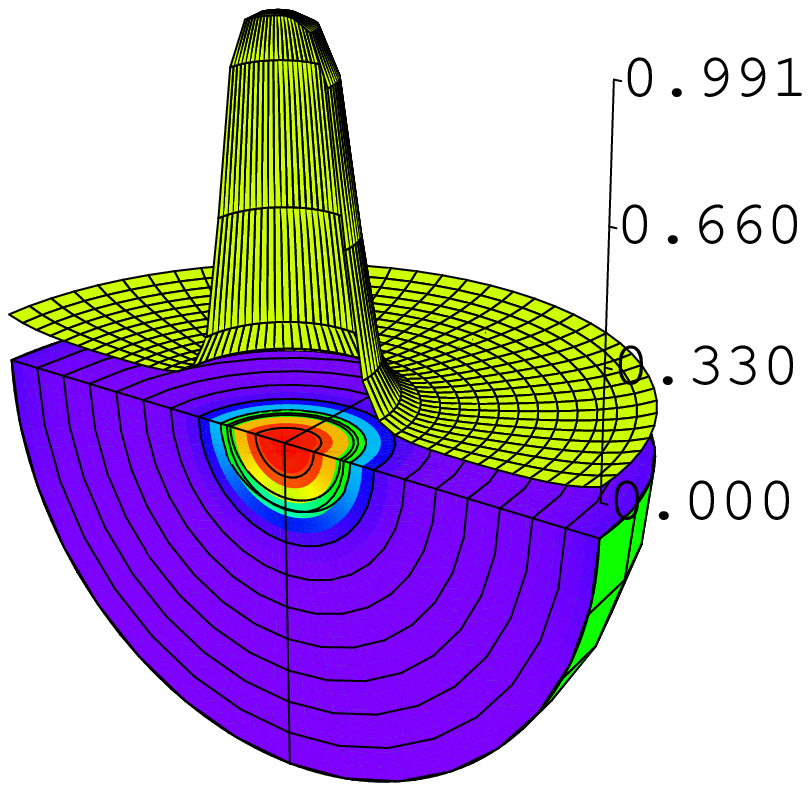}}}
\subfigure[$E=-0.06$,
$L=0.4$]{\scalebox{.49}{\includegraphics{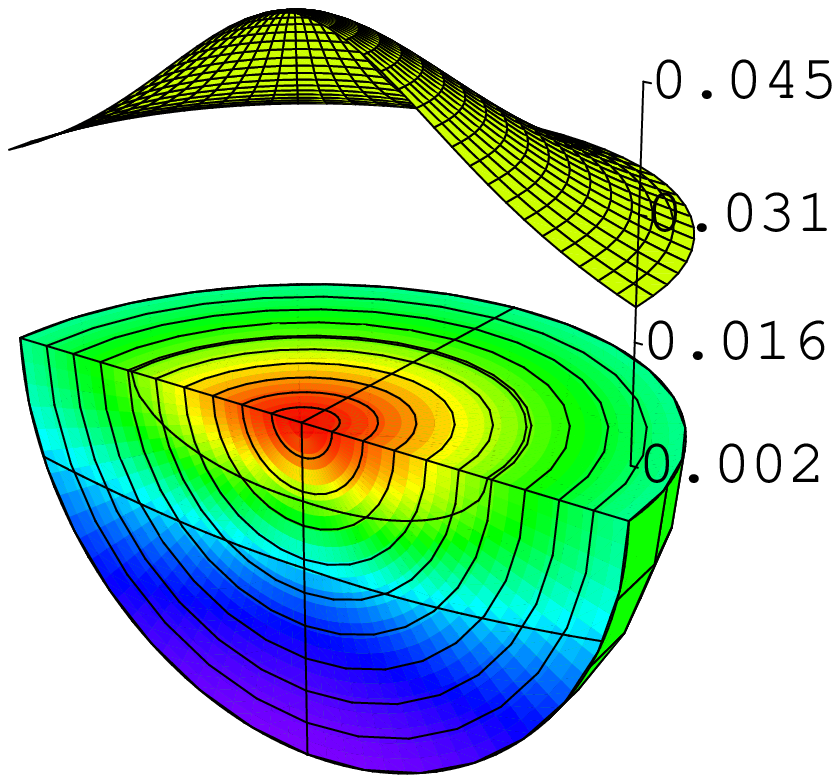}}}
\subfigure[$E=-0.9$,
$L=0.5$]{\scalebox{.47}{\includegraphics{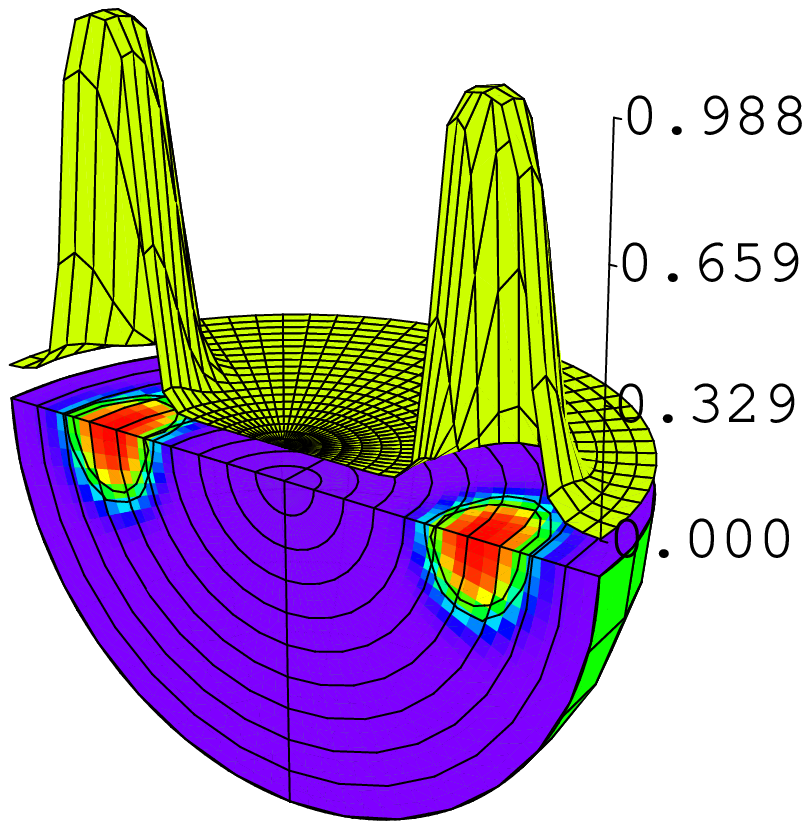}}}
\subfigure[$E=-0.42$,
$L=0.5$]{\scalebox{.47}{\includegraphics{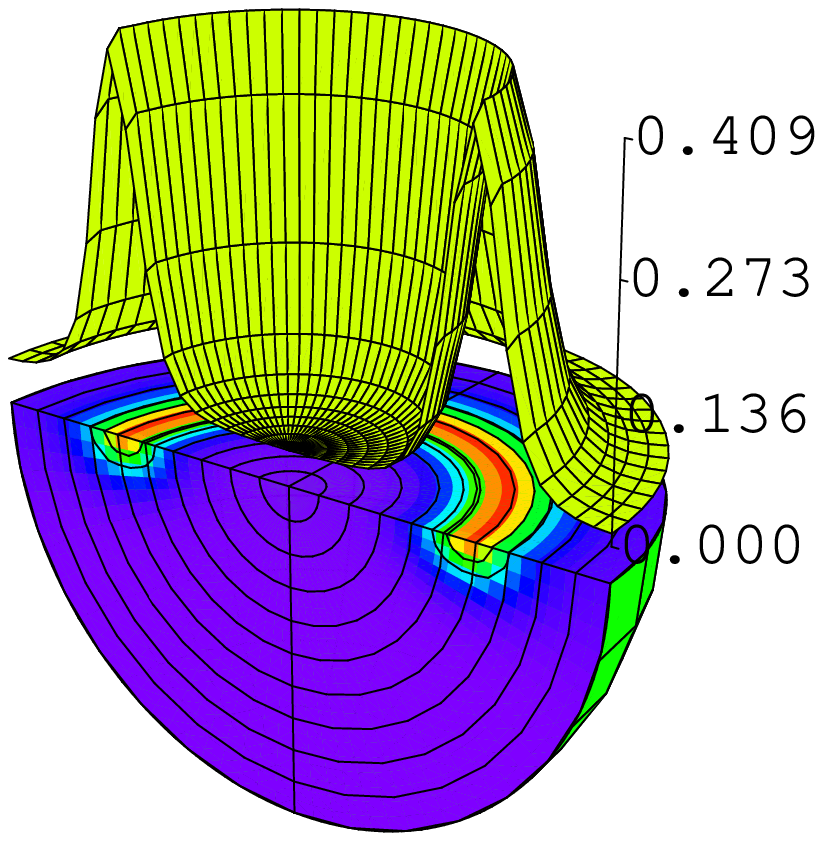}}}
\caption{\label{distri}Examples of equilibrium distributions
$c(\boldsymbol{x})$ occurring inside our spherical box. Shown are
the contour plot and, above it, the density profile. The top row
corresponds to low angular momentum: one sees single clusters
(lower $E$, (a)) and disk-like structures (higher $E$, (b)). The
bottom row displays some results for high angular momentum: double
clusters (lower $E$, (c)) and rings (higher $E$, (d)).
$\boldsymbol{L}$ lies along the vertical axis.}
\end{figure}
The integral-algebraic system (\ref{due}) can be solved
numerically as follows. Fixing $E$ and $L$ and starting from a
reasonable initial guess for $b_{lm}(x)$, one can compute
$u_{lm}(x)$ from (\ref{u}) ($1$-dim. integral). Using this,
(\ref{due}) can be calculated ($2$-dim. integral) to obtain a
better form for $b_{lm}(x)$. This scheme can be iterated until
convergence. Clearly, numerical calculations must be performed
with a finite number of harmonics. A first reduction is obtained
by excluding odd harmonics from the calculation. Exclusion of
$l=1$ fixes the center of mass in the origin, while absence of
higher-order odd harmonics prevents the formation of asymmetric
structures (e.g. two clusters of different sizes lying at
different distances from the origin). Besides this, from the
typical behaviour of $b_{lm}(x)$ obtained from the above
procedure, shown in Fig.~\ref{spag}, one clearly sees that
$b_{lm}$ dies out fast as $l$ increases. Therefore, all
calculations were performed with even harmonics up to and
including $l=16$. We set the particles' masses to one and measured
energy and angular momentum in units of $GN^2/R$ and
$(RGN^3)^{1/2}$, respectively. Solutions of (\ref{key2}) at fixed
$E$ and $L$ were obtained, choosing $\Theta=0.02$ always. For the
sake of simplicity, we took the angular momentum parallel to the
$3$-axis.

The resulting phase diagram is reported in Fig.~\ref{phased}. In
order to discern pure phases from phase-coexistence regions (mixed
phases), the Hessian of $S_N$, namely
\begin{equation}
\mathcal{H}[S_N]=\text{det}
\begin{pmatrix} \partial^2_E S_N & \partial_L\partial_E S_N
\\ \partial_E\partial_L S_N & \partial_L^2 S_N
\end{pmatrix}
\end{equation}
must be analyzed \cite{gross174,gross186}. Indeed, in pure phases
one has $\mathcal{H}[S_N]>0$, while in mixed phases
$\mathcal{H}[S_N]<0$. In the latter, the specific heat is negative
and statistical ensembles (microcanonical and canonical) are
inequivalent. For low angular momenta, the system is more likely
to be found in a single dense cluster (SC) at low energies, while
for higher energies the most probable state is a gaseous cloud
(G). The two phases, both pure, are separated by a mixed phase
with negative specific heat where different equally-probable
equilibrium configurations coexist. One thus recovers the usual
collapse scenario \cite{thirring70} that is found in theories
without angular momentum. For higher angular momenta, instead, the
most probable equilibrium configuration is a double cluster (DC,
pure phase), although the gas remains the most probable at
sufficiently high energies. A sample of equilibrium density
profiles is shown in Fig.~\ref{distri}. A central question is
clearly that of stability of such structures. A detailed analysis
will be given elsewhere, however rings turn out to be unstable, at
odds with single and double clusters.

\begin{figure}
\includegraphics[width=8.5cm,height=6.2cm,clip=true]{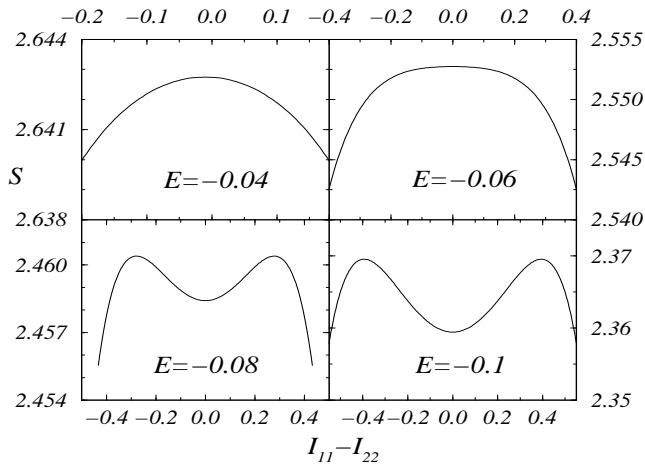}
\caption{\label{rsb}Entropy as a function of the order parameter
$I_{11}-I_{22}$ at $L=0.5$ and different values of $E$. The values
of $E$ and $L$ correspond to the four markers ($\times$) shown in
Fig.~\ref{phased}.}
\end{figure}

The appearance of the previously-unobserved double-cluster
solutions at high angular momenta is particularly remarkable (and
stresses again the importance of angular momentum in
self-gravitating systems), because in such a state rotational
symmetry is spontaneously broken. This is shown explicitly in
Fig.~\ref{rsb}, where the entropy is plotted as a function of the
order parameter $I_{11}-I_{22}$ (measured in units of $NR^2$),
$I_{ab}$ being the components of the inertia tensor $\mathbb{I}$
($a,b=1,2,3$). (We remind the reader that we chose the angular
momentum to be parallel to the $3$-axis.)
If $L=0$ (or in absence of angular momentum) the system is
isotropic ($I_{11}=I_{22}=I_{33}$) and rotational symmetry cannot
be broken. When $L\neq 0$, anisotropies may occur ($I_{33}\neq
I_{11},I_{22}$) and one can have either rotationally-homogeneous
($I_{11}= I_{22}$) or rotationally-heterogeneous ($I_{11}\neq
I_{22}$) solutions. The latter correspond to double clusters.
Indeed, the entropy profile develops two peaks at non-zero values
of $I_{11}-I_{22}$, corresponding to binary-star systems, with the
two stars either aligned on the $1$-axis or on the $2$-axis.

Summarizing, we calculated the static equilibrium density profiles
of a self-gravitating and rotating gas using a microcanonical
mean-field approach, showing that the formation of double-cluster
structures (previously observed only through dynamical approaches)
can be obtained from the spontaneous breakdown of a fundamental
symmetry of the Hamiltonian (\ref{ham}). We have presented for the
first time the global phase diagram as a function of the conserved
quantities $E$ and $L$. The inclusion of angular momentum in this
analysis is absolutely crucial for these results, the formation of
a double-cluster structure being possible only through the
spontaneous breaking of a fundamental symmetry of the Hamiltonian
(i.e. the rotational symmetry). However, it goes by itself that
the formation and stability of stars, double stars, etc. involves
other forces (e.g. nuclear and sub-nuclear) than Newtonian
gravity, and hence could require other ingredients than just
energy and angular momentum \cite{contopoulos60}. The results
presented here provide nevertheless the most detailed static
analysis of the problem to date, and bridge the existing gap
between static and dynamical theories. A more exact picture of the
situation could be obtained by introducing correlations, which are
ignored in the present mean-field approach.

We wish to thank P.-H. Chavanis and O. Fliegans for useful
discussions, comments and suggestions.

\end{document}